*Chapter 1*

# ELECTRIC VEHICLE CHARGING SCHEDULING IN GREEN LOGISTICS: CHALLENGES, APPROACHES AND OPPORTUNITIES

***Luyang Hou, Chun Wang* and Jun Yan***
Gina Cody School of Engineering and Computer Science
Concordia University, Montreal, Canada

## ABSTRACT

Replacing a fossil fuel-powered car with an electric model can halve greenhouse gas emissions over the course of the vehicle's lifetime and reduce the noise pollution in urban areas. In green logistics, a well-scheduled charging ensures an efficient operation of transportation and power systems and, at the same time, provides economical and satisfactory charging services for drivers. This paper presents a taxonomy of current electric vehicle charging scheduling problems in green logistics by analyzing its unique features with some typical use cases, such as space assignment, routing and energy management; discusses the challenges, i.e., the information availability and stakeholders' strategic behaviors that arise in stochastic and decentralized environments; and classifies the existing approaches, as centralized, distributed and decentralized ones, that apply to these challenges. Moreover, we discuss research opportunities in applying market-based mechanisms, which shall be coordinated with stochastic optimization and machine learning, to the decentralized, dynamic and data-driven charging scheduling problems for the management of the future green logistics.

* Corresponding Author Email: chun.wang@concordia.ca

## INTRODUCTION

Logistics has been a key sector in global economies and a crucial contributor to social progress [1]. Both individuals and businesses expect delivery of their goods to be faster, more flexibly, and in the case of consumers, at a lower cost, which poses challenges on logistics transport. Add it all up, the city transportation sector is under intensifying pressure in delivering a better service at an ever-lower cost [1] by taxis, freights, vans, or buses. Transitively, the increasing logistics transport demands will release more greenhouse gas. Therefore, Electric Vehicles (EVs), without tailpipe emissions, are contributing to the fight against localized pollution that is increasingly important in overpopulated urban areas [2].

Governments, e.g., the UK Councils[2], have introduced several measures to curb air pollutions in cities, such as incentivizing EVs, investing in cleaner buses, monitoring borough-wide pollutions, and pioneering the concept of low-emission zones. Some electric logistics projects, such as FREVUE[3] in Europe, provided insights on how innovative solutions using electric freight vehicles could help achieve emission-free city logistics. Companies and cities are required to provide charging service points, including central charge depot and public charging stations, for electric vehicles: decisions need to be made on the number, location, and capacity of these service points constructed.

Unlike fossil fuel-powered vehicles, however, EVs must recharge frequently due to the limited driving range allowed by the battery capacities; worse still, each recharge also takes a significant amount of time. The driving range of EVs for a single charge is around one-third of the petrol-equivalent, while the recharging time can be hours, contrary to the minutes for refueling internal combustion engine vehicles (ICEVs) at a gas station [3]. For logistics, both the frequency and the duration of EV charging are concerning, especially compared to the relatively short customer service times, e.g., small package shippers or a customized pickup, and thus clearly affect the productivity and route planning in logistics [1].

Given these unique features of EVs, green logistics management, on the top of traditional logistics, is emerging as a research field that aims to accommodating different type of EVs, Electric Vehicle Service Equipment (EVSE), and renewable energy resources. Fig. 1 describes the role of EV charging scheduling in green logistics ecosystem that charging scheduling integrates with – and deeply influence – various issues in transportation, power systems, and supply chain management,

[1] The future of the logistics industry- PwC, https://www.pwc.com/transport.

[2] https://techmoneyfit.com/blog/uk-government-invests-in-charging-points-for-electric-taxis/.

[3] Project: Freight Electric Vehicles in Urban Europe, https://frevue.eu/.

etc. It is therefore highly desirable to schedule and coordinate charging activities of EVs in city logistics in order to improve logistics efficiency and reduce the operational costs, due to the following factors:

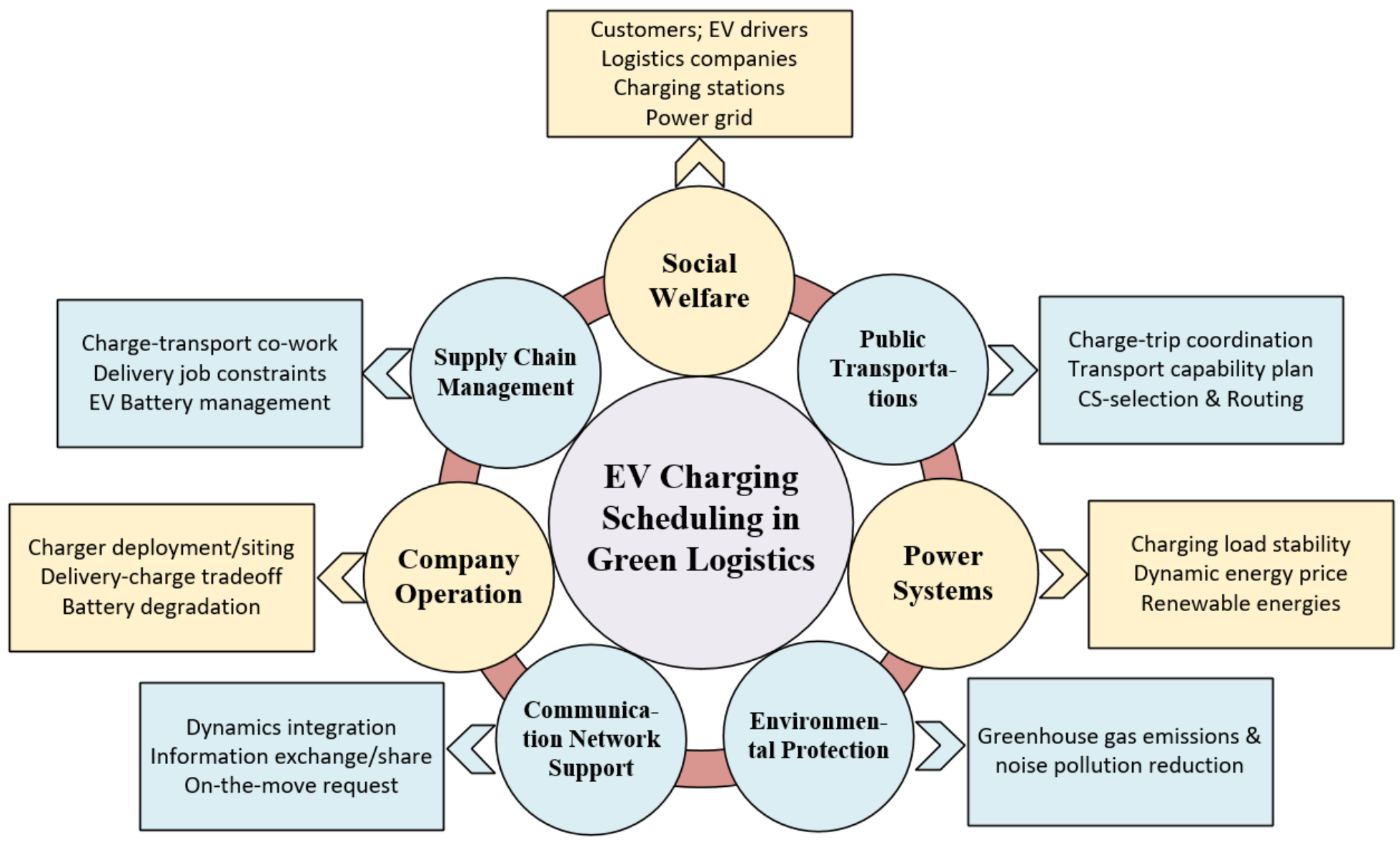

Figure 1: EV charging scheduling in green logistics ecosystem

***Environmental and service incentives***: EVs can greatly reduce the green gas emissions compared to ICEVs when it come to running costs, environmental impact, and quality of driving. Moreover, a well-scheduled charge can reduce drivers' waiting time at charging stations and thus improve drivers' satisfaction [4].

***Co-management of public transportations and power systems***: Scheduling on the charging time or places will influence EVs' service delivery and its route planning in logistics. Moreover, the load induced by EV charging at different charging stations will stress the electricity network that delivers energy to each charging station, and bring negative impacts, i.e., voltage deviation, transformer saturation, power loss and voltage deviation [5] [6] [7]. Thus, charging scheduling should be coordinated with traffic control and energy management, in order to improve transport efficiency and maintain grid stability.

***EV/charging facility ratio***: Currently, the growth of publicly-accessible chargers, especially fast chargers, still falls behind the increase of EVs on the road [8]. The reason could be attributed to the large costs of charging facility investment and long payback period. Given this, charging activities have to be well-scheduled to better utilize the current limited charging capacities, restricted by the number of chargers installed at charging stations and its respective power output, for logistic EVs that often need to meet stringent time requirements.

***Multiple charge requirements***: Logistic EVs are often tasked with heavy transport workloads at various locations throughout a day. It is thus key to schedule delivery jobs or services with multiple recharges in terms of when, where and how much to charge during the trips. A well-designed charging scheduling can increase profits, reduce waiting times, and improve the overall efficiency of both logistics system and charging stations.

***Social welfare***: The efficiency of charging scheduling is highly dependent on the dynamic behaviors and decisions made by stakeholders, i.e., consumers, drivers, and charging service providers, who act as independent, rational yet mostly self-interested agents in a decentralized open environment. Thus, the social welfare is expected to be achieved across all stakeholders in the entire logistics system.

Motivated by these factors, it is of great importance to efficiently coordinate and schedule different charging requests, such as single charge, deferable charge or partial charge, in order to maintain charging load stability, improve drivers and consumer experiences, as well as logistics and transport efficiency. To achieve these goals, different approaches, e.g., mathematical or stochastic optimization, game theory, and machine learning, are imperative to EV charging scheduling problems in the decentralized and stochastic environment so that the social welfare can be maximized in logistics systems.

The remainder of this chapter is structured as follows: Section 2 reviews and analyzes some of the charging scheduling problem in sustainable city logistics and its challenge; Section 3 presents the typical approaches, including centralized, distributed and decentralized approaches, that deals with these challenges in logistical charging scheduling problems; Section 4 provides an overview of potential research opportunities of EV charging scheduling in green logistics.

## PROBLEM FORMULATION AND CHALLENGES

As the research and development of green logistics are evolutionary, we provide a detailed summary and taxonomy of related EV charging scheduling problems. Moreover, our work complements the existing surveys by presenting the challenges for solving these charging scheduling problems in the literature.

## Four-element Structure of EV Charging Scheduling

As a field in operations research, scheduling aims to find the best way to assign the resources to the activities at specific times such that all the constraints are satisfied, and the best objective measures are produced. In spite of the variety of the definitions and models, most scheduling problems can fit in a four-element structure, which consists of resources, jobs, constraints, and objectives [9]. The relationships of these elements are described as: resources are assigned to jobs over the continuous-time or discrete-time[4] and this assignment process is restricted by the constraints and guided by the objectives. Based on this structure, we define the EV charging scheduling as a resource-constrained allocation problem in terms of the following elements:

***Charging requests***: A set of charging activities of electric vehicles that is necessary for accomplishing delivery jobs or services in logistics;

***Resources***: The space and power of charging stations: the space resource refers to the number of installed charging points; the power resources can be distributed generations (photovoltaic system, wind power, hydro turbines, bio-gas, etc.) [10], energy storage system (ESS), and EV battery in Vehicle-to-Grid (V2G) paradigm. A battery swap station can also be treated as an energy resource where applicable;

***Constraint***: A set of conditions that must be satisfied in the charging scheduling process, e.g., precedence constraints, release time and deadlines of request, battery capacity, or the resource capacity constraints. To be specific, constraints can be generally classified into three types: power capacity, limited space (parking space and charging points), and time/energy constraints from users;

***Objective***: A criterion to judge a schedule's performance, which can be classified into two categories: from grid and charging station prospective and from EV users prospective. Users measure the quality of service (QoS), charging costs and their satisfaction. Charging station measures the grid stability and the utilization efficiency of its limited capacity.

## Unique Features

Based on the four-element structure, we categorize logistical EV charging scheduling problems into one classical resource-constrained allocation problem. In contrast to the traditional scheduling problems, EV charging scheduling problem has several unique features in green logistics:

---

[4] In terms of the time representation in scheduling formulations, continuous-time models are potentially allowed to take place at any point in the continuous domain of time. While the whole optimization process in discrete-time models is split into a series of time slots and allocate energy in each time step. The mathematical programs for continuous-time problems are usually of much smaller sizes and require less computational efforts for their solution than the discrete one.

(1) ***Service process in a market***[5]***:*** EV charging can be seen as a service process with the presence of stakeholder-provided information, and these inputs have a significant impact on the output efficiency of systems with which they interact, due to the tight coupling between stakeholders, system operation, and market exchange;

(2) ***Transportation integration:*** Space assignment and routing are crucial transportation problems which decide where and when to activate the charging demand, considering drivers' predefined deadlines, energy demands, charging station availability, and power limits. The objective is to minimize the waiting time, costs, and/or travelling distances by selecting appropriate location and time of charging;

(3) ***Energy management:*** Charging stations need to determine the amount of energy that can be allocated to each plug-in EV during each time period as limited by the power distribution networks and/or energy storage systems. Energy management is extremely important for coordinating the transport and recharge of freights or buses in central charging depots. The energy demand for each trip, timing of each charge, as well as the capacity of each charging station should be considered to provide high-quality services and, at the same time, maintain the power grid stability. Moreover, charging facilities can equipped with intermittent renewable energy sources or distributed generation, such as solar or wind energy, which can pose uncertainties and challenges to charging scheduling;

(4) ***Marginal utility*:** EV users' gain from charging more energy is diminishing/decreasing along the time given the lithium battery charging profile (current decreases at saturation stage of charging curve). In this case, pre-emption is allowed such that a user may adjust or reduce their energy demands to obtain the best time-SoC tradeoff;

(5) ***Battery swap/switch paradigm:*** Battery swapping, as a new energy source instead of charging, could be an efficient and grid-friendly way for EV charging, especially for the frequent transportation works in logistics. The whole operation could take less than ten minutes, which is on par with conventional vehicles and much faster than even some fast recharging stations [1].

## Typical Use Cases and Problems

We analyze some typical use cases and problems of charging scheduling addressed in green logistics, based on the above features in the context of four-element structure, as follows:

---

[5] A market is one of the many varieties of systems, institutions, procedures, social relations and infrastructures whereby parties engage in exchange. It is such complex in economics that we only capture its several important concepts in conducting our research, i.e., competition, individual behaviors, price maker and taker, limited resource allocation, demand and supply, negotiation, and social welfare.

**Vehicle scheduling problem (VSP) with recharge**

VSP models and optimizes vehicle-to-trip assignment problem with EV battery capacity constraints [11]. Each trip across a set of locations has a given time and energy demand that has to be satisfied without exceeding EV's driving range; decisions should be made on when, where and how much to charge a group of EVs while dispatching them to accomplish these trips. The objectives include to maximize the number of tasks that are completed [12], to minimize the number of vehicles used and total distance travelled [11] [13], or to minimize the costs through ahead-of-time charging planning [14]. To achieve these goals, it is key to estimate the energy-related costs and constraints, energy demand and time constraint of each task, and the space and power capacity constraint of charge depot.

A typical application is the scheduling of urban taxis for customers pickup [15]. Taxis have to get sufficient power for the remaining driving distance of next pickup service in order to maximize the total profit, including revenues paid by the passengers, vehicle maintenance costs, vehicle depreciation costs, parking space maintenance costs and parking costs. A well-planned charging can accommodate more electric taxis in urban areas during peak hours, which can ease customer anxiety while improving driver profits.

**Charging scheduling with limited space**

Charging periods or start times are assigned via scheduling to EVs under time constraints (arrival, departure, and charging time) and the limited number of chargers. For instance, J. Timpner and L. Wolf proposed a coordinated charging strategy to integrate the reservation and dynamic charging requests into the charging schedule, in order to improve the utilization of the limited charging places [16]. M. Zhu et al. models the EV charging scheduling problem as one Parallel Machine Scheduling (PMS) problem, which is to schedule EVs to different charging outlets with the total waiting time minimized [17]. The charging time is modelled as a fuzzy number in [18].

To efficiently utilize the limited charging space, a charging cable sharing strategy is proposed to deal with the public charging station coordinated charging [19]. The authors solved optimal configuration of charging stations and scheduling of charging power to each EV during each time interval. The objective is to enhance charging station's utilization level and save corresponding investment costs. Similarly, a charging point sharing paradigm is proposed in [20] to balance the charging space with energy flow at a charging station. The idea is to use an *M* (input)-to-*N* (output) charger, with which the charger output and input are restricted by the limited transformer capacity.

**Routing and charging station selection**

Charging routing problem with energy constraint is to find the most economical route or charging places with the minimum time (wait or driving time) or energy consumption, considering traffic conditions (path planning) and available resources at the charging stations [21] [22]. In addition, the charge station selection problem can be integrated with power allocation [23] [24], which optimizes both transport and charging under the constraint of availability, power capacity, and price of charging stations.

**Multi-aggregator collaboration**

In power community, load dispatching/scheduling coordinates multiple energy demands from EVs to different charging stations in the electric power networks, in order to alleviate the negative effects of charging activities on electric distribution networks [6]. The charging of logistical EVs at central depot should be especially well scheduled due to its heavy burden over the local load. Energy management is an optimal control process for the output power to the plug-in EVs during different time intervals [25] [26], it usually does not consider the limited spaces, but only the limited power capacity in the electric power networks. Vishu Gupta et al. [27] addressed a multi-aggregator-based charge scheduling problem that incorporates collaborative charging and realistic situations with variable energy purchase and cancellation charges. The objective is to maximize the number of EVs that are scheduled at public charging stations, and the total profit of the aggregators.

**Co-management of transportation and power system**

A new research field focuses on the systematic interaction of intelligent transportation and power system during charging scheduling, which considers both of the spatial charging demands from users and the limited space and power capacity of the charging facilities [28]. This problem aims to improve the benefits of resource providers (charging stations) and consumers (EV users).

For instance, Y. Luo et al. [29] proposed a multi-objective charging scheduling strategy for EV charging scheduling and path planning, considered transport and grid related system information, such as road length, vehicle velocity, waiting time, as well as load deviation and node voltage in distribution network. To serve more EV users with random behaviors and demands, H. Chen et al. [30] proposed a two-stage stochastic programming model for charging facility planning, where charging is restricted by limited parking space and power capacity in a multiple-charger multiple-port charging environment. A bi-level smart EV charging scheduling problem in working place is addressed by B. Yagcitekin and M. Uzunoglu in [22], where the first level considers the transformer power demand and transformer capacity from the prospective of power grid, and the second level routes the EVs to the most suitable charging point, and controls the charging process cost-effectively and reliably.

## A Taxonomy for EV Charging Scheduling

In view of above descriptions, we present a taxonomy for the EV charging scheduling problems according to the operational environment, as shown in Table 1. The environmental factors are:

1) **information availability** in an offline/deterministic, stochastic/dynamic, or online/real-time environment;

2) **information asymmetry** in a centralized, distributed, or decentralized environment.

The offline environment assumes all the problem data (e.g., number of jobs, charging times, release dates, due dates, charging facility information, etc.) are known in advance. As for the stochastic environment, distributions of the problem data are known in advance. While in an online environment, central controller does not know the upcoming jobs or charging requests, including number of jobs to be processed, release dates, processing times. Charging requests are presented to the controller one after another in a real-time manner.

In the centralized environment, central controller collects charging requests from the relevant entities and makes the decision to allocate the available resources [51]. The centralized scheme can acquire the optimal solution, in which each entity contributes to the decision-making individually without strategic behaviors and consideration of others' actions.

<table>
<tr><th></th><th></th><th colspan="3">Information Availability</th></tr>
<tr><th></th><th></th><th>Offline/Deterministic</th><th>Stochastic/Dynamic</th><th>Online/Real-time</th></tr>
<tr><td rowspan="4">Information Asymmetry</td><td>Centralized</td><td>[11] [13] [14] [15] [17] [31] [32] [33] [34]</td><td>[16] [18] [35] [36]</td><td>[12] [37]</td></tr>
<tr><td rowspan="2">Distributed</td><td colspan="2">[38] [39] [40] [41]</td><td>×</td></tr>
<tr><td>×</td><td>[19] [21] [28] [42] [43] [44] [45] [46]</td><td>[26] [47] [48]</td></tr>
<tr><td>Decentralized</td><td>[49] [50] [51] [52] [53]</td><td>[25] [54] [55] [56] [57]</td><td>[58] [59] [60] [61]</td></tr>
</table>

Table 1: A Taxonomy of Electric Vehicle Charging Scheduling Problems

The distributed and decentralized environment does not have a control center and allows each entity to make its own decisions in a distributed environment, where entities are autonomous decision makers who are motivated by their own rationalities and not controlled by other entities or a system-wide authority. Moreover, these two schemes both assure scalability, with which scheduling related information is normally scattered across the entities in the system, and no entity has a global view of the problem [9].

However, decentralized setting assumes that the self-interested entities may strategically misrepresent their energy consuming patterns and preferences to the resource providers, such that they can receive more monetary gains. This strategic manipulation may bring unexpected outcomes and low efficiency. But the favorable point lays on entity privacy protection, which prevents central authorities from collecting information in decision making [62].

## Challenges for EV Charging Scheduling

This part analyzes the complexities derived from applying current approaches to EV charging scheduling problems in green logistics. Basically, charging scheduling in green logistics arise two level of challenges: from the basic scheduling domain and from the environment factors.

The scheduling domain complexity involved in solving the most NP-hard scheduling problems is the central theme of EV charging scheduling, which is related to the computational requirements to generate an outcome given EV users' charging requests and different sorts of constraints.

Aside from the computational complexity, information availability and strategic behaviors due to information asymmetry pose additional challenges on the top of traditional scheduling domain complexity in solving EV charging scheduling in logistics. These challenges will be further amplified when involving and coordinating large number of entities. Below are the details on the complexity derived from the environmental factors:

### Information availability in stochastic and online environment

The charging activities in green logistics are always operated in highly distributed and dynamic environments, with uncertainties coming from charging time [18] [31], availability of charging stations [28], arrivals of EVs [31, 59, 63], and energy prices [47] [63]. Different sorts of uncertainties exist in practical scenarios where some data are not precisely known. These uncertainties will influence the decisions made on when, where, and how much to charge an EV based on the target of optimization procedure. Sometimes a sequence of decisions, rather than a single decision, need to be made, e.g., taxis with several partial charges during the day trip; these decisions often depend crucially on the dynamic aspects of the environment.

Moreover, uncertainties also exist in the power systems, such as the state of the electricity grid, the production of renewable sources, the charging point availability, the congestion at communication and transportation networks, and the number of EVs available to provide V2G services, which are changing quickly while a large number of EVs are either driving or charging [7]. Therefore, maintaining load stability is especially challenging in Micro-grids[6] due to the uncertainties from renewable energy supplies and the lower load capacity compared to power grid.

**Strategic behaviors in decentralized environment**

In a decentralized environment, stakeholders are not cooperative but strategic, and charging resource allocation can be viewed as distributed optimization problems, with an objective function that depends on the strategic behaviors and private information of the stakeholders in the system. Typical examples of decentralized charging scheduling problem include task allocation across multiple self-interested shipping companies and decisions made by electric drivers on the trade-off between recharge and jobs pickup. The presence of stakeholder inputs is a necessary and sufficient condition to define a charging scheduling as a service process.

Since the stakeholders are self-interested agents who aim to maximize its own utility, the challenges that significantly affect the social welfare of the whole society can be attributed to four factors from agents' standpoint:

1) agents may be reluctant to participate in the scheduling process;

2) agents may misrepresent their energy demands and preferences on charging pattern, such as deadlines, charging time, and energy requirements;

3) agents may be stubborn or insensitive to alter their charging or electricity consuming habits to gain greater benefits in response to market signals;

4) agents may be unaware of the precise representation of their valuations or preferences.

In terms of these characteristics, market-based mechanisms can be a natural way to tackle the strategic behaviors and information asymmetries in decentralized logistics systems. Typical market mechanisms refer to game theory and auction in games, which captures the conflicting economic interests of the resource providers and resource consumers [49, 52, 55, 58]. Users can interact and negotiate with each other via information exchange, and further, coordinate their electricity usage with others to achieve a social welfare.

[6] A Micro-grid is a small-scale power production and delivery system comprising distributed generation facilities co-located with the loads they serve.

In addition to the incentive mechanisms in game-theoretic design, Time-of-Use (TOU), dynamic pricing and float pricing in DR program are also efficient strategies in providing incentives for EV users to change their charging habits as their best response relied on the other users' economic rationality, by allowing collectives of users to participate in the charging resource allocation [62]. The prices are retained fixed within different pricing periods ahead of time. Users receive this signal and are motivated to reduce or shift their power demands and energy usage by observing these price signals in a competitive market. However, these pricing schemes do not involve those utility theory and strategic behaviors (such as misreport information) of users in a decentralized environment, as the information revealed by users is guaranteed to be truthful. It is always challenging to elicit users' true preferences over the charging and sensitivity on the changing of time and energy price.

## EXISTING MODELLING PARADIGMS AND APPROACHES

In this section, we will review the most recent activities relevant to the optimization of charging scheduling problems in logistics. Existing works typically use either mathematical programming or utility-based agent coordination combined with the mechanism design approach, such as auction and game theory, to model the charging scheduling problems in dealing with the complexities of information availability and asymmetry. The scheduling approaches in literature can be generally classified into centralized, distributed and decentralized based approach with respect to the aforementioned challenges. A classification of existing approaches is provided in Table 2.

| | Offline/Deterministic | Stochastic/Dynamic | Online/Real-time |
|---|---|---|---|
| Centralized | Mathematical optimization; Meta-heuristics | Robust optimization | Meta-heuristics; Machine learning |
| Distributed | Mathematical optimization; Meta-heuristics | Stochastic optimization; Distributionally robust optimization; Pricing strategy | Online optimization; Markov decision process |
| Decentralized | Game theory; Combinatorial auction | Auctions - reinforcement learning modelling; Stackelberg game | Online mechanism design |

Table 2: Existing approaches for the operational challenges

## Centralized/Deterministic Modelling Paradigm

Theoretically, a centralized/deterministic modelling paradigm allows for achieving the best solution as the central authority has access to all information about charging scheduling. However, the difficulty of this approach lies in application bottlenecks such as scalability, computation tractability, data privacy concerns and communication infrastructure [64]. The model parameters (energy demand, arrival times, charging time, etc.) in centralized/deterministic environment are known with certainty even though they are only the estimations of values in real-world applications. Moreover, it provides best solutions and constitutes the basis for solving a dynamic and distributed charging scheduling problem, as witnessed by various successful applications in logistical EV charging scheduling problems [14, 17, 31, 32, 34].

### Mathematical optimization

To model and solve EV charging scheduling problems, extensive works apply linear programming (LP) [36], dynamic programming (DP) [65], mixed integer linear programming (MILP) [15] [28]; decomposition techniques: Lagrangian Relaxation (LR) [63], Lagrangian decomposition (LD) [66], or robust optimization [25] [28] and stochastic programming [38] [39] in developing a variety sorts of scheduling algorithms or strategies. These approaches follow a *centralized* scheme under the coordination by a central controller with known parameters. One advantageous aspect for the scheduling results is the optimal solution and highest efficiency can be obtained with the complete and truthful information from users.

### Meta-heuristic optimization

Meta-heuristic methods, such as Genetic Algorithms (GA) [18], Particle Swarm Optimization (PSO) [67], and Artificial Bee Colony Algorithm (ABC) [68], etc., can efficiently explore large search spaces and incorporate heuristic knowledge on the problem domain in NP-hard problems. A meta-heuristic provides a sufficiently good solution to the charging scheduling problem, especially with incomplete or imperfect information or limited computation capacity.

Although meta-heuristics do not guarantee global optimal solutions compared to the exact algorithms, they are still playing an important role in charging scheduling and routing problems in a centralized environment. For instance, a two-phase heuristic algorithm is also used in routing planning of taxis in [13] to minimize the total travel distance. The nearest-neighbor heuristic adds the closest customer to extend a route in the first phase, and two types of move operations, exchange and relocate, are combined with a Tabu search in the second phase.

## Modelling Paradigm with Uncertainty/Dynamics

In this part we will review some typical modelling paradigms for EV charging scheduling with uncertainties in stochastic/dynamic environments, in which the parameters are known only in probability; that is, random variables for which a probability distribution of possible parameter realizations are known, but the variability of possible values must be modelled to choose the best values for the decision variables in the optimization.

### Stochastic optimization

*Stochastic programming (SP):*

SP is a powerful modeling framework for optimization problem under uncertainty which optimizes the expected objective value across all the uncertainty realizations. SP can accommodate decision making process with single/two/multi-stages. The most widely used one is two-stage model, where deterministic optimization problems with known parameters are formulated in the first stage, while the stochastic problems with some unknown parameters (however its probability distribution is known) are modelled in the second stage [47] [48]. SP is efficient for modelling two-stage energy management in Demand Response programs, i.e., day-ahead demand scheduling and real-time power control, the goal is to improve the efficiency, reliability and safety of the power system, through motivating changes in the customers' power consumption habits [48] [62].

For instance, a two-stage model by SP is proposed in [38], where the energy scheduling with the day-ahead power market is solved in the first stage, and the real-time energy scheduling is solved in the second stage. The objective is to find solutions that are feasible for all possible scenarios while minimizing the expected cost at the first stage. Similar works refer to [39, 40, 41].

*Robust Optimization (RO) and Distributionally Robust Optimization (DRO):*

RO is recently introduced to model uncertainties in charging scheduling, such as renewable energy supplies, energy prices, and drivers' energy demands. This approach is suitable for situations where the range of the uncertainty is known but the distribution of uncertainty is not [25]. While stochastic programming assumes there is a probabilistic description of the uncertainty, robust optimization works with a deterministic, set-based description of the uncertainty, which constructs a solution that is feasible for any realization of the uncertainty in a given set.

Moreover, data-driven optimization under uncertainty requires distributionally robust optimization [25] [28], also known as data-driven stochastic program, where the uncertainty is modeled by a set of probability distributions, namely ambiguity set. DRO can obtain prior knowledge of the probability distributions through historical and/or real-time data, in terms of the practical scenarios where the precise information of the ambiguity set is rarely available or known. For instance, the day-ahead energy management model incorporated uncertain market prices using RO and used stochastic optimization to model the uncertain charging demand (arrival, departure, and charging times of EVs at charging station) [25]. An uncertainty set is constructed for market prices to minimize the mismatch of the realized specific prices and the forecast one, and thus may decrease charging stations' monetary losses. Moreover, a data-driven robust optimization model is developed in [28] to optimize the capacities of renewable generations and energy storage units in each charging station, where the output uncertainty of photovoltaic energy and charging demands are formulated as robust chance constraints.

*Markov Decision process (MDP):*

Some discrete-time stochastic charging scheduling problem can be modelled as an MDP process, with the typical time-driven scheduling policy adopted. An MDP model is typically defined as a 5-tuple: 1) decision epoch; 2) action; 3) state; 4) transition probability; and 5) reward and cost functions. MDP can investigate the constrained stochastic optimization problem in terms of the uncertainty, for instance, the arrival of EVs, the intermittent renewable energy, or the variation of the energy price [63]. If the probabilities or rewards are unknown, the problem is one of reinforcement learning in practical deployment [69] [70]. Typical work refers to D. C. Parkes et al. [59], which modelled the online mechanism design problem as an MDP to solve an energy unit allocation problem. The optimal policies are implemented in a truth-revealing Bayesian-Nash equilibrium.

**Online optimization**

Some charging scheduling problems adopt a model-free online scheme, where uncertain data are collected sequentially in a real-time fashion. The time horizon is slotted in equal intervals in time-driven mode and scheduling decisions are made at each time interval. For instance, a distributed offline and online framework is proposed in [26] to collaborate multiple aggregators for scheduling, in order to maximize the total profit of the aggregators and the total number of vehicles charged. However, F. Kong et al. point out that the major dilemma for applying the time-driven policy for charging scheduling is to determine the length of time slots [61]. Long time slots lead to few charging mode switches but cause under-utilized charging points at the stations, while short time slots improve charging point utilization but cause many mode transitions for EVs. Given this, event-driven could be an efficient solution for online charging scheduling.

**Machine learning based approach**

Data-driven optimization under a highly stochastic and distributed environment that integrates machine learning and mathematical programming is appealing in the era of big data, which can predict EV mobility, charging demands, load fluctuation, renewable energy generation, as well as other system uncertain parameters in green logistics. For example, Artificial Neural Network (ANN) is applied in the optimal energy management for a day-ahead price forecasting, so that the error between the actual and predicted electricity prices and the cost of parking lot owner with respect to the time of use can be minimized [65]. Similarly, ANN with sample average approximation is used for predicting the base load power consumption [38].

Some charging scheduling problem are modelled as MDP and solved by reinforcement learning, which is a model-free algorithm that can directly integrate user feedback into its learning process [71]. For instance, the cost-effective day-ahead consumption plan can be learned to better forecast numerous details about each EV behavior (e.g., plug-in times, power limitations, battery size, power curve, etc.) [70].

**Queueing theory**

Queueing theory addresses routing and charging station selection issue, in order to find the most appropriate charging sites with minimum waiting time and balance the traffic flow among different stations [21, 42, 72]. This distributed scheduling is used to assign multiple charges to different charging stations, which is often applied to the highway scenario. For instance, S. Bae and A. Kwasinski [73] proposed a spatial and temporal model of electric vehicle charging demand, which first predicts arriving rate of EVs by the fluid dynamic traffic model, and then forecasts the charging demand by queueing theory.

## Modelling Paradigm with Strategic Behaviors

To deal with EV drivers' strategic behaviors in a decentralized environment, general equilibrium theory in game theory [49, 50, 55] and mechanism design in micro-economic theories [52, 58, 61] incentivize users to participate in the scheduling process, represent their real charging preferences, and alter their charging or electricity consuming habits to gain greater benefits. These market-based mechanisms are widely applied for energy until allocation and aggregator collaboration in either offline [49] [55] or online environment [59, 60, 61].

The equilibrium in game-theoretic models is defined as the condition that each participant acts on its best-response strategy with respect to others' strategy and cannot benefit itself by unilaterally deviating from this current state with an alternative strategy [74]. For instance, the energy exchange process is modelled as a non-cooperative Stackelberg game in [50], in which the smart grid acts as a leader, who needs to decide on its price so as to optimize its revenue; while the EVs act as followers, who need to decide on their charging strategies so as to optimize the trade-off between the benefit from battery charging and the associated cost. A distributed algorithm enables the EVs and the smart grid to reach a generalized Nash equilibrium (GNE). In addition, a cake cutting game is applied in [54] to deal with the selection of EVs and route for transportation demands, in which the limited idle time for the serving EVs should be efficiently utilized for charging. The goal is to balance the transportation and charging demands to guarantee the long-term operation of photovoltaic systems with less charging costs and more profits.

The most important mechanism-design application in market setting is auctions. In a decentralized environment, users can negotiate with the electricity network on the power allocation at different time intervals, through mechanism design based-approaches [53] [59]. For example, a pricing process for multi-tenancy autonomous vehicle servicing problem is modeled as a combinatorial auction based on Vickrey-Clarke-Groves (VCG)-based charging mechanism in [52], in which the service providers, as bidders, compete for offering transportation services; as a result, the social welfare is maximized. Moreover, a type of Groves mechanisms is proposed in [51] to allocates the available charging capacity (discrete energy unit) under network constraints at the distribution networks, this mechanism is able to obtain a Nash equilibrium and is shown to be efficient and strategy-proof.

## RESEARCH OPPORTUNITIES

Multi-agent systems (MAS) architecture provides a natural modeling of the distributed and stochastic aspects of charging markets, the existing agent-based simulation platforms from both academic and commercial sectors will provide invaluable tools for validating emerging models and techniques for solving EV charging scheduling problems. Stakeholders, e.g., electric freights/taxis/buses, consumers, charging stations, logistics companies, distribution network operators and energy generators, can be modelled as strategic, rational and self-interested *agents* in the context of decentralized system engineering.

In this section, we will discuss several research challenges and opportunities in designing market-based mechanisms for addressing the charging scheduling issues in decentralized and stochastic environments. Under the multi-agent systems architecture, future research is expected to go towards automated coordination systems for EV charging by applying mechanism design, game theory, stochastic optimization, and machine learning based approaches, in the implementation of real-time coordination and forecasting methods that can be used by the agents to adjust its forthcoming operation and properly schedule their charging.

### Grid-interactive Transportation: X + Charging Scheduling

EV charging scheduling is playing a crucial role in the systematic interaction among logistics systems, intelligent transportation systems and smart grids, as its performance greatly impacts public transport efficiency, logistics costs, charging load stability, as well as agents' satisfaction. Different charge patterns, such as single/multi-charge, deferrable charge, or partial charge with adjustable energy demands, will introduce uncertainties and flexibilities to EV charging scheduling. Add it up, charging scheduling will arise various interesting issues in the integration with *highway travel/transport*, *ride-sharing*, *self-driving paradigm*, *e-taxi/Uber service*, *smart manufacturing transport*, *path planning*, *V2G/V2B/ V2V*, *mobility on demand (MoD) scheme*, etc. A systematic efficient outcome considering agents' transport needs with energy demands and the limited charging capacities is supposed to be obtained, in order to maximize the social welfare and resource utilization level.

### Decentralized, Dynamic, Data-driven: A 3D Prospective

The interactions between stakeholders, privacy maintenance from user side, uncertainties and dynamics in charging market are the key dimensions of charging scheduling problems in logistics. Dynamic information about drivers' charging requests and charging stations' availability have been extensively researched in a dynamic and decentralized environment. For instance, in terms of an aggregated EV charging scheduling problem with energy storage, an offering/bidding strategy of an ensemble of charging stations coupled in the day-ahead electricity market is proposed in [25], where aggregator determines optimal bidding strategy for the amount of energy to sell and buy from the market to meet the aggregated demands. The uncertainty modelling of the market price used robust optimization, and aggregated charging station demand used stochastic optimization.

However, big data and strategic behaviors pose additional challenges to the optimization with uncertainty. And market mechanism design should also deal with the changing of private information over time based on dynamic auction [75]. Two typical works refer to [76] [77]: an optimal auction paradigm with deep learning is proposed in [76], where the rules of an auction are modelled as a neural network, and machine learning is used for the automated design of auctions with budget constraints. Moreover, a closed-loop data-driven optimization framework is discussed in [77], where a loss function that incorporates the objective function of mathematical programming could be used to train the machine learning model, and feedback from the model-based system serves as input to the data-driven system.

Several advantageous properties of model-based system compared to model-free system include strong performance guarantees and explainable outcomes. However, these techniques often do not assure scalability and may not be applicable in problems for which *a priori* information is unavailable [78]. In light of this, online learning methods could be solvable, but large-scale mechanism design problems in conjunction with big data faces computational challenge for the training of machine learning and winner determination.

To our best knowledge, these is no such an effective mechanism for tackling decentralized, dynamic, and data-driven EV charging scheduling problem. Game theoretic-based auction design with machine learning in a large-scale environment, on the top of incentive mechanisms and stochastic optimization techniques, are promising areas in future logistics.

### Market-based Mechanisms: Competition to Cooperation

Game theory, stochastic optimization and machine learning have been extremely successful in tackling a wide variety of problems in transportation system, logistics system, and smart grid [39, 40, 50, 58, 61]. Currently, considerable body of literature in game theoretic-based auction design explores *competition* between agents for the limited charging resources, such as energy allocation or charge reservation. Various incentive policies in a competitive market, either Groves mechanisms or dynamic pricing, encourage agents to express their true and complete preferences, or modify their habits in response to market signals, to achieve a systematic efficiency.

However, agents are encouraged, as not only a price-taker, but also a price-maker, to join the resource allocation actively and pursue their benefits that are aligned with the social good. L. J. Ratliff et al. [78] pointed out in the presence of competition, markets could get stuck at a bad equilibrium where all agents play myopic strategies without performing sufficient exploration for a social-desirable outcome. Moreover, their truthful revelation about preferences should not be a necessity for the decision-making because they may not be able to develop a specified valuation model and the valuation may change over the course of the auction.

In terms of this, a simultaneous ascending auction based auction, i.e., combinatorial clock auction [79], could be a potential solution, where the price discovery can help bidders to form their true valuations, alter their demands and yield the resources to more-needed ones, thus ultimately leads to an equilibrium with an efficient assignment, such as, in ride-sharing.

We believe that market-based mechanism design is going towards *cooperation* between self-interested agents, such that the information asymmetries, exogenous uncertainties from dynamic environments, endogenous uncertainties from agents' preferences, and the resource constraints can be well addressed in green logistics.